\documentclass[twocolumn,onecolappendix,trackchanges]{aastex6}

\usepackage{framed}
\usepackage{amsmath}

\usepackage{comment}

\slugcomment{Accepted to PASP}
\shorttitle{Total Power Map to Visibilities (TP2VIS)}
\shortauthors{Koda et al.}

\begin{document}

\title{Total Power Map to Visibilities (TP2VIS): \\ Joint Deconvolution of ALMA 12m, 7m \& Total Power Array Data}

\author{Jin Koda\altaffilmark{1,2,3},
Peter Teuben\altaffilmark{4},
Tsuyoshi Sawada\altaffilmark{2,3},
Adele Plunkett\altaffilmark{5,6},
Ed Fomalont\altaffilmark{6,3}}
\altaffiltext{1}{Department of Physics and Astronomy, Stony Brook University, Stony Brook, NY 11794-3800, USA}
\altaffiltext{2}{NAOJ Chile Observatory, National Astronomical Observatory of Japan,  Alonso de C\'{o}rdova 3788, Office 61B Vitacura, Santiago, Chile}
\altaffiltext{3}{Joint ALMA Observatory, Alonso de C\'{o}rdova 3107, Vitacura, Santiago 763-0355, Chile}
\altaffiltext{4}{Department of Astronomy, University of Maryland, College Park, MD 20742-2421, USA}
\altaffiltext{5}{European Southern Observatory, Alonso de C\'{o}rdova 3107, Vitacura, Santiago 763-0355, Chile}
\altaffiltext{6}{National Radio Astronomy Observatory, 520 Edgemont Road. Charlottesville, VA 22903, USA}
\email{jin.koda@stonybrook.edu}

\begin{abstract}
We present a new package for joint deconvolution of ALMA 12m, 7m, and Total Power (TP) data,
dubbed ``Total Power Map to Visibilities (\textsc{Tp2vis})".
It converts a TP (single-dish) map into visibilities on the \textsc{Casa} platform,
which can be input into deconvolvers (e.g., \textsc{Clean}) along with 12m and 7m visibilities.
The \textsc{Tp2vis} procedure is based primarily on the one discussed in \citet{Koda:2011nx}.
A manual is presented in the \textsc{Github} repository (\url{https://github.com/tp2vis/distribute}).
Combining data from the different ALMA arrays is a driver for a number of science topics,
namely those that probe size scales of extended and compact structures simultaneously
(e.g., protostar outflows and environment, AGB stars and planetary nebulae, molecular clouds and cores,
and molecular clouds in galaxies).
We test \textsc{Tp2vis} using model images, one with a single Gaussian 
and another that mimics the internal structures of giant molecular clouds.
The result shows that the better $uv$ coverage with \textsc{Tp2vis} visibilities clearly helps
the deconvolution process and reproduces the model image within errors of only 5\% over
two orders of magnitude in flux.
In the Appendix, we describe how the model image is generated.

\end{abstract}

\section{Introduction}

The Atacama Large Millimeter/submillimeter Array (ALMA) consists of three sets of telescopes:
two interferometers with 12m and 7m dishes and Total Power (TP) 12m single-dish telescopes.
The interferometers alone cannot obtain information on an extended emission distribution.
Combined, these three sets can recover structures from small to large scales.
This emphasizes the importance of single-dish data in the imaging process.
Interferometers collect Fourier components of a 2-dimensional brightness distribution in the sky,
but cannot obtain full coverage of the $uv$ space (i.e., Fourier space).
Hence, the empty part of the $uv$ space has to be filled (or ``guessed")
in the deconvolution process, with the filled part of the $uv$ space used as a guideline.
Convergence is not guaranteed, but a better sampled $uv$ space helps this process.
The combination of interferometer and single-dish data provides better $uv$ coverage,
thus leading to better imaging results.

The idea of combining interferometry and single-dish data was introduced early
\citep[who called it ``short-spacing corrections"]{Bajaja:1979aa, Ekers:1979aa},
but in practice the procedure is still under debate.
The history and techniques for combination are summarized in \citet{Stanimirovic:2002aa}.
The techniques can be classified into three general categories:
whether the data are combined (1) {\it before}, (2) {\it during}, or (3) {\it after} deconvolution (e.g.,  {\sc Clean}).
For better $uv$ coverage for deconvolution,
(1) and (2) have natural advantages \citep[e.g., ][]{Cornwell:1988aa}.
Each of these categories can be split into two subcategories:
whether the operations are performed (a) in the Fourier domain ($uv$ domain) or (b) in the image domain.

\subsection{Combination After Deconvolution}

A common technique of the after-deconvolution combination (category 3a)
applies low- and high-pass filters to single-dish and {\it cleaned}
interferometer maps (or images), respectively.
The maps are then added in the Fourier domain and transformed back to a final map.
It is now called the Feather technique.
Several implementations exist
\citep[e.g., the {\sc Imerge} task in {\sc Aips}; {\sc Immerge} task in {\sc Miriad}; ][]{Herbstmeier:1996aa, Weis:2001hq}.
The {\sc Feather} task\footnote{\url{https://casa.nrao.edu/casadocs/latest/global-task-list/task_feather/about}}
in {\sc Casa} is one more recent example.
It adopts the low-pass filter, $L(u,v)=FT\{B_{\rm TP}(l,m)\}$, and a high-pass filter, $H(u,v)=1-L(u,v)$,
where the Fourier transform (FT) of the TP primary beam $B_{\rm TP}(l,m)$ is used
\footnote{{\sc Feather} first deconvolves a TP map with $B_{\rm TP}$, and then
multiplies $L(u,v)=FT\{B_{\rm TP}(l,m)\}$ in the Fourier domain.
Therefore, it recovers the original TP map. In other words, {\sc Feather} simply adds the original TP map.}.
{\sc Feather}'s choice of $H(u,v)$ extends to $uv\approx 0$ which interferometer data does not cover.
To avoid this, \citet{Blagrave:2017aa} used different filters with non-zero values only where the data exist.
Since multiplications (or divisions) in the Fourier domain are equivalent to convolutions
(deconvolutions) in the image domain, it is possible to implement the same technique
both in the Fourier and image domains.
\citet{Faridani:2018aa} implemented the {\sc Feather} technique in the image domain (category 3b).

\subsection{Combination During Deconvolution}

The TP map can be used as an initial model for deconvolution of interferometer data (category 2a).
Practically, such a scheme deconvolves single-dish and interferometer data jointly \citep{Cornwell:1988aa}.
The {\sc Clean}/{\sc Tclean} tasks in {\sc Casa} can take a single-dish map as an input.
\citet{Dirienzo:2015aa} employed this technique and produced a combined map.
They further applied {\sc Feather} to ensure that the large-scale components
precisely represented the TP data.

\subsection{Combination Before Deconvolution}

Pseudo visibilities can be generated from a single-dish map.
They can be added to interferometer visibilities before deconvolution \citep[category 1a; ][]{Vogel:1984aa}.
Once TP visibilities are generated from the TP map,
they are readily fed into existing deconvolvers (e.g., {\sc Clean}/{\sc Tclean}).
This technique has also been used, though there are some discussions on 
how to optimize the distribution and weights of the pseudo visibilities
\citep[e.g., ][]{Rodoriguez-Fernandez:2008rf, Kurono:2009aa, Pety:2010aa, Koda:2011nx}.
For example, \citet{Koda:2011nx} generated a Gaussian visibility distribution and set
the visibility weight so that the single-dish primary beam is represented.
One of the advantages of pseudo visibilities techniques is
that the visibility weight can be easily controlled within single-dish data,
as well as with the interferometer data.

The same technique can be implemented in the image domain \citep[category 1b; ][]{Stanimirovic:1999aa}.
The single-dish and interferometer dirty beams, as well as their maps, are added
with a choice of $L(u,v)$ and $H(u,v)$.
The combined dirty map can then be deconvolved with the combined dirty beam. 
This is equivalent to the pseudo visibility techniques,
since a FT of a sum of two functions is equal to the sum of the FT of the two functions.
The weights can be changed by adjusting $L(u,v)$ and $H(u,v)$.
Such implementation is being developed in {\sc Casa} (U. Rau in preperation).

\subsection{This Work}

Tools are needed for the combination of ALMA data
in the platform of the Common Astronomy Software Applications \citep[{\sc Casa}; ][]{McMullin:2007aa}.
Here, we introduce a new package in {\sc Casa} that converts ALMA's TP map into pseudo visibilities (dubbed {\sc Tp2vis}).
It generates visibilities as if they were observed by a virtual interferometer
with short spacings down to zero spacial frequency.
Together with ALMA12m+7m data, the {\sc Tp2vis} visibilities can be fed
into standard deconvolvers for joint deconvolution.
The technique follows the one developed by \cite{Koda:2011nx} on the {\sc Miriad}
platform \citep{Sault:1995kl}, but with improvements.

The \textsc{Tp2vis} package provides four functions in the \textsc{CASA} platform.
The key function, \textsc{Tp2vis}, generates a \textsc{CASA} measurement set (MS)
--visibilities with weights-- from a TP map.
The weights are calculated based on the root-mean-square (RMS) noise from the TP map.
This MS is ready for joint deconvolution, though in some cases one may wish
to manipulate the weights using the supplementary function, \textsc{Tp2viswt}.
An accessory function, \textsc{tp2vispl}, plots the weights of ALMA 12m, 7m,
and TP visibilities for comparison.
Another function, \textsc{tp2vistweak}, attempts to fix
the problem of beam-size mismatch in the image space after deconvolution
(Section \ref{sec:beamproblem}).

\begin{deluxetable*}{clc}
\tablecaption{Beam Terminology\label{tab:beamterm}}
\tablewidth{0pt}
\tablehead{
\colhead{Number} & \colhead{Beam Name}  & \colhead{Symbol}}
\startdata
1 & Primary beam of Total Power (TP) telescopes  & $\Omega_{\rm TP}$ \\
 \cline{1-3}
2 & Primary beam of Virtual Interferometer (VI) telescopes & $\Omega_{\rm vir}$ \\
 \cline{1-3}
3 & Dirty beam (output from clean/tclean with niter=0) & $\Omega_{\rm dirty}$ \\
& Synthesized beam & $\Omega_{\rm syn}$ \\
& Deconvolution beam & $\Omega_{\rm dec}$ \\
& Point spread function & $\Omega_{\rm psf}$ \\
 \cline{1-3}
4 & CLEAN beam  (typically derived from Gaussian fit to $\Omega_{\rm dirty}$) & $\Omega_{\rm clean}$ \\
& Convolution beam & $\Omega_{\rm conv}$ \\
& Restoring beam & $\Omega_{\rm res}$
 \enddata
\tablecomments{Horizontal lines separate different beams.
The third and forth beams have a few names:
$\Omega_{\rm dirty}=\Omega_{\rm syn}=\Omega_{\rm dec}=\Omega_{\rm psf}$, and
$\Omega_{\rm clean}=\Omega_{\rm conv}=\Omega_{\rm res}$.}
\end{deluxetable*}

\section{The TP2VIS procedure} \label{sec:steps}

\begin{figure*}
\epsscale{0.8}
\plotone{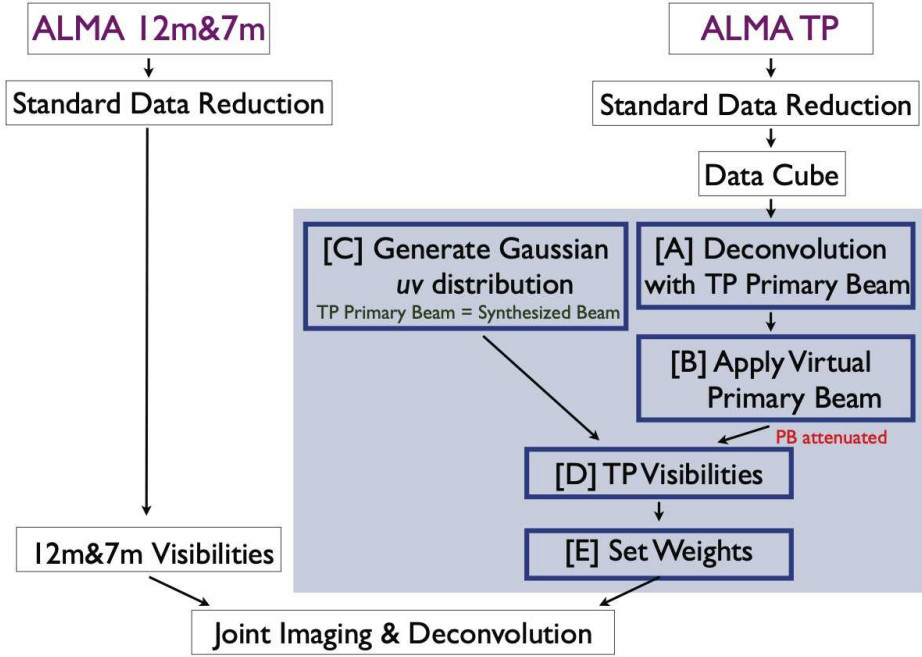}
\caption{Flowchart of the \textsc{Tp2vis} procedure.
\textsc{Tp2vis} processes the operations shaded in blue: taking a calibrated TP data cube
from \textsc{CASA}, converting it into a set of visibilities (i.e., a measurement set),
and passing it to a deconvolution task in  \textsc{CASA}.
The TP visibilities from \textsc{Tp2vis} can be jointly deconvolved
with the ALMA 12m and 7m interferometer visibilities.
\label{fig:flowchart}}
\end{figure*}

The \textsc{Tp2vis} function takes a sky brightness distribution (TP map)
and simulates observations with a {\it virtual} interferometer (defined below).
Figure \ref{fig:flowchart} shows the flowchart of steps in the \textsc{tp2vis} function.
The details of the steps are in the following subsections.
From a top view, the procedure can be separated into three parts.
First, it converts the TP map into the sky brightness distribution,
which will be observed by the virtual interferometer (steps A and B in the figure).
Second, it converts the brightness distribution into visibilities (steps C and D).
And third, the weights of the TP visibilities are set so that they represent
the RMS noise of the original TP map (step E).

Different types of beams are involved in our discussions.
For clarity we list the beams in Table \ref{tab:beamterm}.

\subsection{Step A: Deconvolution with TP Beam $\Omega_{\rm TP}$}
This step is a preparation for the virtual interferometer observations.
When the TP telescopes observe the sky, the TP telescope beam,
$\Omega_{\rm TP}$, is convolved with the brightness distribution in the sky.
Therefore, the first step is to deconvolve the TP map with $\Omega_{\rm TP}$
and to obtain the true sky brightness distribution in Jy/beam.

We approximate $\Omega_{\rm TP}$ with a Gaussian with a full width half
maximum (FWHM) of $56.6\arcsec$ at 115.2 GHz.
We assume that it scales linearly with 1/frequency.

\subsection{Step B: Apply Virtual Primary Beam $\Omega_{\rm vir}$}
When an interferometer observes the sky, the brightness distribution is
attenuated by its primary beam (PB).
Hence, the sky brightness distribution from Step A is multiplied by
the PB pattern of the virtual interferometer ($\Omega_{\rm vir}$).
In case of mosaic observations, this must be done at each pointing position.

In our virtual observations, the coordinates of pointing centers can be
set arbitrarily within the TP map.
We typically take them from the ALMA 12m data,
so that the 12m and TP visibilities have the same sky coverage.
The pointing coordinates are listed in an ASCII file and passed to
the \textsc{Tp2vis} function with the ``ptg=" argument.

\subsection{Step C: Generate Gaussian Visibility Distribution}
In parallel with steps A and B,  a visibility distribution for the virtual observations
(hence a MS) is prepared. We generate a two-dimensional Gaussian distribution of
visibilities in the $uv$ space, so that its Fourier transformation,
i.e., the synthesized beam $\Omega_{\rm syn}$, reproduces the Gaussian
beam of the TP telescopes ($\Omega_{\rm TP}$).
Adopting a constant weight for all TP visibilities (Section \ref{sec:weights}),
this distribution would represent the sensitivity distribution within the TP beam.
The width of the Gaussian visibility distribution is determined from
those of the Gaussian beam.
The standard deviations of the Gaussian beam in the sky $\sigma$
and the Gaussian distribution in the $uv$ space $\sigma_{\rm F}$ are related as
$\sigma_{\rm F}=1/(2\pi\sigma)$.

The full width at half maximum (FWHM) is calculated as
FWHM=$2\sqrt{2 \ln 2}\sigma$ ($\approx 2.355\sigma$).
For a FWHM of $56.6\arcsec$, $\sigma=2.74\times 10^{-4}\,\rm radian$
and $\sigma_{\rm F}=580\lambda$, where $\lambda$ is the wavelength.

\subsection{Step D: Fill Visibility Amplitudes and Phases}
We already have the sky brightness distribution with the PB of the virtual interferometer
applied (from step B), and the Gaussian visibility distribution in the $uv$ space (step C).
Now, the sky brightness distribution is Fourier transformed into the $uv$ space.
The amplitude and phase of the transformation are then sampled at the visibility positions,
which fill the AMPLITUDE and PHASE columns of the visibilities in the MS.

These visibilities are already {\it internally} consistent.
The \textsc{Clean/Tclean} tasks with natural weighting produce
a synthesized beam $\Omega_{\rm syn}$ roughly equal to the TP telescope beam $\Omega_{\rm TP}$.
The brightness distribution also reproduces the observed TP brightness distribution.

\subsection{Step E: Set Weights of TP Visibilities}

The \textsc{Tp2vis} function sets visibility weights according to the RMS noise in the TP map.
The RMS value is set manually by the ``rms=" argument [in units of Jy/beam of the TP map].
[Note that in the current implementation, the RMS is set by hand,
because it is not always trivial to find emission-free channels using an algorithm.]
The RMS is converted to the weight of individual visibilities.
The equation for this conversion is in Section \ref{sec:weights}.
The sensitivity-based weight is a proper representation of the quality of the TP data
and is the default of \textsc{Tp2vis}.

In the history of radio interferometry, the weights are often manipulated.
The most common examples are the uniform, robust/briggs, or taper weighting schemes.
There is no reason not to manipulate them in some other ways -- for example,
to match the weights of TP visibilities with those of ALMA 12m and 7m data
(their sensitivities may not match when one works on archival data).
The supplementary \textsc{tp2viswt} function provides
several options to manipulate the weights.
An accessory \textsc{tp2vispl} function plots the weight distributions of ALMA 12m, 7m,
and TP visibilities for comparison, so that users can decide how the TP visibilities
need to be weighted with respect to 12m and 7m data.

\section{Beam Disparity and Flux Problem}

Imaging of interferometer data involves several different beams.
Some of them are often referred with different names by different researchers or in different circumstances.
This is very confusing.
Furthermore, their differences cause a more practical issue in flux conservation
in the process of imaging.
In order to discuss the issue, we first explain the beams.
For \textsc{Tp2vis} we use four different beams (Table \ref{tab:beamterm}).

The first two beams were used in Section \ref{sec:steps}:
the primary beam of the TP telescopes ($\Omega_{\rm TP}$)
and that of the virtual interferometer ($\Omega_{\rm vir}$).
The third and fourth beams are from visibilities,
i.e., a synthesized beam ($\Omega_{\rm syn}$) and
convolution beam ($\Omega_{\rm conv}$).
The $\Omega_{\rm syn}$ is also called the dirty beam ($\Omega_{\rm dirty}$),
deconvolution beam ($\Omega_{\rm dec}$), or point spread function ($\Omega_{\rm psf}$).
The $\Omega_{\rm conv}$ is also called the clean beam ($\Omega_{\rm clean}$)
or restoring beam ($\Omega_{\rm res}$).
For discussions in later sections, we give more precise definitions of
$\Omega_{\rm syn}$ and $\Omega_{\rm conv}$ in Section \ref{sec:synconvbeam}.

\citet{Jorsater:1995aa} extensively discussed the problem of flux conservation
when there is a disparity between the areas of $\Omega_{\rm syn}$ and $\Omega_{\rm conv}$.
The problem becomes more apparent when the TP visibilities are included
(Section \ref{sec:beamproblem}).
We discuss two techniques to resolve/mitigate this problem:
by adjusting the weights of the TP visibilities (Section \ref{sec:wtbeam})
and/or by scaling the residual map (Section \ref{sec:imagetweak}).

Note that we use the symbols, $\Omega_{\rm syn}$ and $\Omega_{\rm conv}$,
to represent the synthesized and convolution beams, respectively,
but also to denote their areas.

\subsection{The Synthesized and Convolution Beams} \label{sec:synconvbeam}

The synthesized beam $\Omega_{\rm syn}$
is determined by the distributions of visibility weights in $uv$ space
[Note: $\Omega_{\rm syn}=\Omega_{\rm dirty}=\Omega_{\rm dec}$].
Its 2-dimensional pattern, $B(l,m)$, is a Fourier transformation
of the weight distribution $W(u,v)$.
With the normalization of $\iint W(u,v)dudv=1$,
\begin{equation}
B(l,m) = \iint W(u,v) e^{+2\pi i(ul+vm)}dudv.\label{eq:synbeamb}
\end{equation}
The area of $\Omega_{\rm syn}$ is the beam solid angle
and is given as an integration over space,
\begin{equation}
\Omega_{\rm syn}= \iint B(l,m) dl dm = W(0,0). \label{eq:omegasyn}
\end{equation}
It depends only on $W(0,0)$, the weight at ($u$,$v$)=(0,0).
Therefore, only the weight of the TP visibility matters, because
the interferometer data have zero weight at ($u$,$v$)=(0,0).
Therefore, $\Omega_{\rm syn}=W_{\rm TP}(0,0)$
when ALMA 12m, 7m, and TP are combined.

The convolution beam $\Omega_{\rm conv}$
is typically defined by a Gaussian fit
to the central peak in the $B(l,m)$ pattern
[Note: $\Omega_{\rm conv}=\Omega_{\rm clean}=\Omega_{\rm res}$].
Hence, it depends on the full weight distribution, $W(u,v)$;
but in practice, the outer, long-baseline part of the visibility weight distribution
is often the determining factor.

In summary, $\Omega_{\rm syn}$ is set at the center of
the $uv$ space, while $\Omega_{\rm conv}$ is determined mainly by the outer part.

\subsection{Beam Disparity and Flux Problem} \label{sec:beamproblem}
For simplicity, we use \textsc{Clean} as an example here,
but it applies to other deconvolvers as well.
The \textsc{Clean} task iteratively finds peaks in a dirty map, whose
brightness units are Jy/$\Omega_{\rm syn}$.
In each iteration \textsc{Clean} identifies a peak in a dirty map,
fits the 2-dimensional pattern of $\Omega_{\rm syn}$ to the peak
and obtains its position and amplitude,
subtracts it from the dirty map,
and puts a $\delta$-function at that position in a clean-component map.
The \textsc{Clean} task then convolves the clean-component map with
$\Omega_{\rm conv}$ to produce a model map.
The model map has brightness units of Jy/$\Omega_{\rm conv}$.
Typically, \textsc{Clean} is not perfect and leaves emission in a residual map.
The unit of the residual map remains Jy/$\Omega_{\rm syn}$.
\textsc{Clean} {\it naively} adds the model and residual maps to produce a final map,
whose unit is often denoted as Jy/beam.
For the ``beam" in the denominator, $\Omega_{\rm conv}$ is typically adopted.
Therefore,
\begin{equation}
\left[\frac{F}{({\rm Jy}/\Omega_{\rm conv})} \right] = \left[\frac{M}{({\rm Jy}/\Omega_{\rm conv})} \right] + \left[\frac{R}{({\rm Jy}/\Omega_{\rm syn})}, \right].\label{eq:finalmap}
\end{equation}
where $F$, $M$, and $R$ stand for the final, model, and residual maps, respectively.

The areas and patterns of $\Omega_{\rm syn}$ and $\Omega_{\rm conv}$ could be very different,
as they depend on different parts of the weight distribution in the $uv$ space (Section \ref{sec:synconvbeam}).
In particular, if their areas are not equal, $\Omega_{\rm syn} \neq \Omega_{\rm conv}$,
the brightness of $R$ in eq. (\ref{eq:finalmap}) is calculated incorrectly.
For example, if $\Omega_{\rm syn} \gg \Omega_{\rm conv}$,
the brightness in $R$ is overestimated by $\Omega_{\rm syn} / \Omega_{\rm conv}$ ($ \gg 1$),
leading to an overestimation of flux in $F$.
This happens when the ALMA 12m, 7m, and TP visibilities are combined.
In theory, the $R$ could be zero if we could clean down to zero flux,
but in practice, it is nearly impossible to get there, as the map to be cleaned has noise.

In case of $\Omega_{\rm syn} \ll \Omega_{\rm conv}$,
the brightness in $R$ is underestimated.
The impact of this on $F$ may not be as noticeable as in the opposite case,
since $R$ typically has much less flux than $M$.
This happens when the 12m and 7m visibilities are imaged without TP,
since $\Omega_{\rm syn}=W_{\rm TP}(0,0)=0$ (hence, $\ll \Omega_{\rm conv}$).
This flux problem did not attract much attention in the past,
as single-dish data were typically not included.

\section{Weights}\label{sec:weights}

The optimal weighting scheme for TP visibilities is a topic of debate.
\textsc{Tp2vis} provides some implementations of weighting schemes, as well as
providing a framework for testing other schemes in the future.
It has been very common to manipulate visibility weights in the history of radio
interferometry.  For example, the natural weighting scheme takes the sensitivities
of individual visibilities measured during observations, and uses them as
a weight distribution in the $uv$ space.
The uniform weighting scheme takes a binary approach, setting the weights to 1
at the positions in the $uv$ space where visibilities exist and 0 for the rest,
independent of sensitivities.
The robust/briggs weighting connects the natural and uniform weightings and transforms
one from the other seamlessly by adjusting the single ``robust" parameter.
Of course, one could come up with other arbitrary weighting schemes to accomplish
particular scientific goals.
No matter what weights are used, the final map should be usable for
scientific applications, as long as the same weights are used consistently for both map
and beam by deconvolvers.

Our main function, \textsc{Tp2vis}, sets the weights so that their sum represents the RMS
noise of the original TP map.
This is the default in \textsc{Tp2vis}.
In case that users wish to manipulate the weights in other ways,
the supplementary function, \textsc{Tp2viswt}, provides options
to manipulate the weights. Currently, it has five modes:

\begin{itemize}
\item {\it mode='statistics'}: 
output statistics of the current weights in MS(s).

\item {\it mode='constant', value=}:
set the weights of all visibilities to a constant specified by ``value=".

\item {\it mode='multiply', value=}:
multiply the current weights by a constant specified by "value=".

\item {\it mode='rms', value=}:
use the sensitivity-based weight for all visibilities (Section \ref{sec:wtsen}).
An RMS value from TP map should be set with ``value=".
This is the default weights in \textsc{Tp2vis}.

\item {\it mode='beam'}:
use the beam size-based weight for all the {\it TP} visibilities (Section \ref{sec:wtbeam}).

\end{itemize}

In what follows, we discuss two provisional approaches to optimizing the weight,
one based on the RMS noise (Section \ref{sec:wtsen})
and another that matches the areas of synthesized and convolution beams (Section \ref{sec:wtbeam}).
This beam matching is important to ensure flux conservation in deconvolution,
and we will revisit this in Section \ref{sec:imagetweak}.
\textsc{Tp2vis} adjusts the weight in the $uv$ space, but the same can be done in the image domain (Section \ref{sec:wtnotes}).

\subsection{The Sensitivity-Based Weight}\label{sec:wtsen}

One of the most intuitive weighting schemes is the one
that represents the quality of TP data,
namely the root-mean-square (RMS) noise of the TP map.
This is the default weighting scheme used by the \textsc{Tp2vis} function,
and the \textsc{Tp2viswt} function with the mode='rms' can set
the weights to the sensitivity-based one.
This section explains the calculation of the sensitivity-based weights in \textsc{Tp2vis}.
Section \ref{sec:wtsentp} discusses the conversion of the RMS noise of a TP map
into the weights of individual visibilities.
For a joint deconvolution, the weights of 12m, 7m, and TP data should be set consistently.
Hence, section \ref{sec:wtsenalma} shows the corresponding weights of 12m and 7m data.
[Note that \textsc{CASA} version 5 uses this definition of 12m and 7m weights by default,
improving upon prior CASA versions.]

This sensitivity-based weighting is equivalent to the conventional ``natural" weighting.
For interferometer data, additional manipulations, such as ``robust/briggs" or ``uniform" weighting
schemes, are often applied on top of the sensitivity-based weights.
Such manipulations can also be applied to the sensitivity-based weights
from \textsc{Tp2vis}. For simplicity, however, the discussions in this section will focus
on the sensitivity-based (natural) weights.

\subsubsection{Weights for TP Visibilities}\label{sec:wtsentp}

With the natural weighting, the sensitivity of a map,
or the RMS noise of the image ($\Delta S^{i}$), is a simple summation over
all visibility sensitivities ($\Delta S^{v}_{k}$),
\begin{equation}
\left( \frac{1}{\Delta S^{i}} \right)^2 = \sum_{k} \left( \frac{1}{\Delta S^{v}_{k}} \right)^2 = N_{\rm vis} \left( \frac{1}{\Delta S^{v}} \right)^2,
\end{equation}
where $N_{\rm vis}$ is the number of visibilities.
All visibilities are assumed to have the same sensitivity ($\Delta S^{v}_{k}=\Delta S^{v}$),
and a single RMS value is assumed for the map.
Hence, the weight of each TP visibility is
\begin{equation}
w_k \equiv \left( \frac{1}{\Delta S^{v}_{k}} \right)^2=\frac{1}{N_{\rm vis}} \left( \frac{1}{\Delta S^{i}} \right)^2.\label{eq:wttp}
\end{equation}

\subsubsection{Corresponding Weights for 12m \& 7m Visibilities}\label{sec:wtsenalma}

For consistency, the 12m and 7m data should also use the sensitivity-based weights.
The sensitivity for a single visibility $k$ between antennas $i$ and $j$ is given by
\begin{equation}
\Delta S^{v}_{k} = C_{ij} \sqrt{\frac{T_{{\rm sys}, i} T_{{\rm sys}, j}}{B \cdot t_{\rm vis}}} =C_{ij} \sqrt{\frac{T_{{\rm sys}}^2 }{B \cdot t_{\rm vis}}},\label{eq:intvissen}
\end{equation}
where $T_{{\rm sys}, i}$, $T_{{\rm sys}, j}$, and $B$, $t_{\rm vis}$ are
the system temperatures of $i$ and $j$, bandwidth, and integration time of the visibility, respectively
\citep{Taylor:1999ab, Thompson:2007fk}.
For simplicity, we assume $T_{{\rm sys}, i}=T_{{\rm sys}, j}=T_{\rm sys}$.
The coefficient $C_{ij}$ is 
\begin{equation}
C_{ij} = \frac{2 k_B}{\sqrt{(\eta_{a,i} A_{i})(\eta_{a,j} A_{j})}} \frac{1}{\sqrt{2} \eta_q},\label{eq:intcoef}
\end{equation}
where $\eta_{a,i}$, $\eta_{a,j}$, $A_{i}$, $A_{j}$, and $\eta_q$ are the aperture efficiencies and
areas of antennas $i$ and $j$, and the quantum efficiency of correlator, respectively.
With the natural weighting, the RMS sensitivity of final map is
\begin{equation}
\Delta S^{i} = C_{ij} \sqrt{\frac{T_{{\rm sys}}^2 }{B \cdot t_{\rm tot}}}, \label{eq:intimgsen}
\end{equation}
where $t_{\rm tot}$ is the total integration time $t_{\rm tot}=N_{\rm vis} t_{\rm vis}$.
The weights of 12m and 7m visibilities should be set to $w_k=(1/\Delta S^{v}_{k})^2$ with eq. (\ref{eq:intvissen}).

\subsubsection{For Internal Consistency}

This subsection is not necessary in practice, but is included for consistency among all data columns in the MS.
It might become important in future versions of \textsc{CASA}.
\textsc{Tp2vis} fills the WEIGHT and SIGMA columns. Only the WEIGHT
column is used by \textsc{Clean}/\textsc{Tclean} in \textsc{Casa} -- currently, other columns do not matter.
However, one could also make the EXPOSURE column (i.e., integration time of each visibility) consistent.

The sensitivity equations for TP, corresponding to
eqs. (\ref{eq:intvissen}), (\ref{eq:intimgsen}), and (\ref{eq:intcoef}), are
\begin{equation}
\Delta S^{v}_{k} = C_{\rm TP} \sqrt{\frac{T_{{\rm sys}}^2 }{B \cdot t_{\rm vis}}}\label{eq:sinvissen}
\end{equation}
and
\begin{equation}
\Delta S^{i} = C_{\rm TP} \sqrt{\frac{T_{{\rm sys}}^2 }{B \cdot t_{\rm tot}}},\label{eq:sinimgsen}
\end{equation}
where
\begin{equation}
C_{\rm TP} = \frac{k_B}{ \eta_a A} \frac{1}{\eta_{\rm mb}} \frac{1}{\eta_q},
\end{equation}
and $\eta_{\rm mb}$ is the main beam efficiency of the TP antennas.
$T_{\rm sys}$, $\Delta S^{i}$, and $B$ are determined by observation.
With eq. (\ref{eq:sinimgsen}) we can derive $t_{\rm tot}$, using $T_{\rm sys}$ and $\Delta S^{i}$
from MS and TP maps, respectively.
Equations (\ref{eq:wttp}), (\ref{eq:sinvissen}), and (\ref{eq:sinimgsen}) give $t_{\rm vis}=t_{\rm tot}/N_{\rm vis}$.

\subsection{The Beam Size-based Weight} \label{sec:wtbeam}
Section \ref{sec:beamproblem} discussed 
the problem in flux due to the disparity between the areas of $\Omega_{\rm syn}$
and $\Omega_{\rm conv}$.
The beam size-based weighting scheme discussed here is an attempt to adjust
the weights of TP visibilities to equalize their areas.
Note again that we use the symbols, $\Omega_{\rm syn}$ and $\Omega_{\rm conv}$,
to represent the areas of the beams as well as to refer the beams themselves.

\subsubsection{The Synthesized and Convolution Beams from Combined INT+TP Visibilities}

The pattern of the synthesized beam $\Omega_{\rm syn}$, denoted by $B(l,m)$,
is given by eq. (\ref{eq:synbeamb}).
The weight is normalized as $\iint W(u,v)dudv=1$.
For interferometers (hereafter INT; e.g., ALMA 12m and 7m) and TP,
the synthesized beam patterns, $B_{\rm INT}(u,v)$ and $B_{\rm TP}(u,v)$, are derived 
with their normalized weights, $W_{\rm INT}(u,v)$ and $W_{\rm TP}(u,v)$, respectively.

By defining a relative weight parameter $\beta$, a combined weight distribution of
INT and TP can be expressed as
\begin{eqnarray}
W(u,v) &=& \frac{W_{\rm INT}(u,v)+\beta W_{\rm TP}(u,v)}{\iint [W_{\rm INT}(u',v')+\beta W_{\rm TP}(u',v')]du'dv'} \\
&=& \frac{1}{1+\beta} \left[ W_{\rm INT}(u,v)+\beta W_{\rm TP}(u,v) \right]. \label{eq:fullwt}
\end{eqnarray}
The corresponding synthesized beam pattern is
\begin{equation}
B(l,m) = \frac{1}{1+\beta} \left[ B_{\rm INT}(l,m) + \beta B_{\rm TP}(l,m) \right].\label{eq:synbeamb2}
\end{equation}

The area of the synthesized beam is derived from eqs. (\ref{eq:omegasyn})(\ref{eq:fullwt}),
\begin{equation}
\Omega_{\rm syn} = W(0,0) = \frac{\beta }{1+\beta} W_{\rm TP}(0,0),\label{eq:synbeam}
\end{equation}
where we used $W_{\rm INT}(0,0)=0$ since INT has no contribution at zero spacing.
Note that $W_{\rm TP}(0,0)$ is equal to the area of the synthesized beam of TP visibilities.

The convolution beam $\Omega_{\rm conv}$ is typically derived by a 2-d Gaussian fit to
the central peak in $B(l,m)$.
Using the major and minor axis diameters, $b_{\rm maj}$ and $b_{\rm min}$,
from the fit, the area of $\Omega_{\rm conv}$ is
\begin{equation}
\Omega_{\rm conv} = \frac{\pi b_{\rm maj} \times b_{\rm min}}{4 \ln 2}.\label{eq:decbeam}
\end{equation}

\subsubsection{Setting Up Weights for TP Visibilities}

This weighting scheme attempts to achieve $\Omega_{\rm syn} = \Omega_{\rm conv}$.
Equation (\ref{eq:synbeam}) gives $\beta$ as
\begin{equation}
\beta = \frac{\Omega_{\rm syn} }{W_{\rm TP}(0,0) - \Omega_{\rm syn} } = \frac{\Omega_{\rm conv} }{W_{\rm TP}(0,0) - \Omega_{\rm conv} }. \label{eq:beta}
\end{equation}
$W_{\rm TP}(0,0)$ is equal to the area of the synthesized beam of TP visibilities alone, which,
by construction, is the same as that of the Gaussian primary beam of the TP antennas.
Hence, 
\begin{equation}
W_{\rm TP}(0,0) = \frac{\pi b_{\rm TP}^2}{4 \ln 2},
\label{eq:wtp00}
\end{equation}
using the FWHM beam size of the TP antennas, $b_{\rm TP}$.
The $\Omega_{\rm conv}$ should be calculated from INT+TP visibilities, but in practice
it depends primarily --and almost solely-- on the longest baselines.
Hence, for simplicity, we derive $\Omega_{\rm conv}$ from INT visibilities alone.
We can calculate $\beta$ from eqs. (\ref{eq:beta}), (\ref{eq:wtp00}), and (\ref{eq:decbeam}) for INT.

Using $\beta$, we adjust the TP visibility weight distribution from $W_{\rm TP}$ to $\beta W_{\rm TP}$,
by scaling the weights of individual visibilities
($w^{\rm INT}_k$ and $w^{\rm TP}_k$ for INT and TP visibilities in MSs, respectively).

For the following, we use the notation $\overline{W}$ for
an {\it un}normalized weight distribution.
The ratio of the two terms in the numerator of eq. (\ref{eq:fullwt}) is
\begin{equation}
\begin{split}
W&_{\rm INT}(u,v):\beta W_{\rm TP}(u,v)  \\
&=\frac{\overline{W}_{\rm INT}(u,v)}{\iint \overline{W}_{\rm INT}(u,v)dudv}:\beta W_{\rm TP}(u,v) \\
&= \overline{W}_{\rm INT}(u,v): \beta \left[ \iint \overline{W}_{\rm INT}(u,v)dudv  \right] W_{\rm TP}(u,v).
\end{split}
\end{equation}

This transformation converts the normalized $W_{\rm INT}(u,v)$ into the unnormalized $\overline{W}_{\rm INT}(u,v)$,
which is what is delivered from the ALMA observatory.
The term to the right of the ratio mark gives the corresponding normalization for the weights of TP visibilities.
When the natural weighting is used,
the unnormalized $\overline{W}_{\rm TP}(u,v)$ and $w^{\rm TP}_k$ are related as
$\iint \overline{W}_{\rm TP}(u,v)dudv = \sum_k w^{\rm TP}_k =N_{\rm vis}^{\rm TP} w^{\rm TP}_k$.
Thus, for the normalized $W_{\rm TP}(u,v)$, we set $w^{\rm TP}_k = 1/N^{\rm TP}_{\rm vis}$.
Also, the term in the bracket is simply,
$\iint \overline{W}_{\rm INT}(u,v)dudv = \sum_k  w^{\rm INT}_k$.
Therefore,
\begin{equation}
w^{\rm TP}_k  = \frac{\beta}{N^{\rm TP}_{\rm vis}} \sum_k w^{\rm INT}_k.
\end{equation}
This weight would satisfy $\Omega_{\rm syn} = \Omega_{\rm conv}$ when the INT and TP visibilities
are combined.

\subsection{Notes on Visibility and Weight Distributions}\label{sec:wtnotes}

The weight distribution of TP visibilities can be controlled,
either by changing the distribution of TP visibilities in $uv$ space,
and/or by adjusting the weights of individual visibilities.
\textsc{Tp2vis} takes the former approach: generating a Gaussian visibility distribution
with a constant weight for all the visibilities.

The latter approach could be useful in future.
For example, one could generates a grid distribution,
instead of the Gaussian distribution, and change the visibility weights as a function
of $uv$ distance so that the weight distribution follows a Gaussian.
This way, one could avoid shot noise due to the randomly-distributed discrete visibilities.
In the end, the \textsc{Clean} task would put the visibilities onto a grid.
If the grid spacing is controlled consistently between \textsc{Tp2vis} and \textsc{Clean},
we can test grid-based approaches in future.
Of course, the weight distribution can be realized on a grid without visibilities.
An implementation of this would require an update of the \textsc{Clean}/\textsc{Tclean} task in \textsc{Casa}
(U. Rau 2018, in private communication).

\section{The Residual Scaling} \label{sec:imagetweak}

The flux problem in Section \ref{sec:beamproblem} is due to the
disparity between the areas of $\Omega_{\rm syn}$ and $\Omega_{\rm conv}$ in eq. (\ref{eq:finalmap}).
The residual part of the final image from \textsc{Clean} is added incorrectly with inconsistent units of brightness.
To circumvent this inconsistency, we could rescale the residual map by a factor of
$\Omega_{\rm conv}/\Omega_{\rm syn}$ and add it to the model map to
generate a new final map.
The \textsc{tp2vistweak} function does this operation
\footnote{One should keep it in mind that the patterns of $\Omega_{\rm syn}$ and
$\Omega_{\rm conv}$ are still not the same, even though their areas are the same.}.

In theory, $\Omega_{\rm conv}/\Omega_{\rm syn}$ can be calculated with
the images of $\Omega_{\rm syn}$ from \textsc{Clean}, but it turns out not to be that simple.
The estimation of $\Omega_{\rm syn}=W(0,0)$ depends on how the visibilities are gridded
in the imaging process.
The integration of the synthesized beam, $\Omega_{\rm syn}=\int B(l,m)dldm$,
requires an unrealistically large image size
as its tiny responses at the far outskirts add up and their area is cumulatively large.

Instead, \cite{Jorsater:1995aa} suggested a more practical approach by taking advantage
of detected emissions.
In parallel to eq. (\ref{eq:finalmap}), we can write a similar equation for the dirty map $D$,
\begin{equation}
\left[\frac{\rm D}{({\rm Jy}/\Omega_{\rm syn})} \right] = \left[\frac{\rm I}{({\rm Jy}/\Omega_{\rm syn})} \right] + \left[\frac{\rm R}{({\rm Jy}/\Omega_{\rm syn})} \right],\label{eq:imaging2}
\end{equation}
where $I$ is a clean-component map, i.e., 
 a map of the emissions that will be identified by \textsc{Clean}.
We should note that the brightness unit is Jy/$\Omega_{\rm syn}$ for all terms,
and that $M$ and $I$ in eqs. (\ref{eq:finalmap}) and (\ref{eq:imaging2})
represent exactly the same emissions in different units.
The $D$, $F$, and $R$ are given by \textsc{Clean}.
We can calculate the scaling factor as
\begin{equation}
\frac{\Omega_{\rm conv}}{\Omega_{\rm syn}} = \frac{F-R}{D-R}. \label{eq:beamratio}
\end{equation}
We multiply the $R$ map by this factor and add it to $M$ to construct a new $F$ as in eq. (\ref{eq:finalmap}),
but with a consistent brightness unit (consistent beam areas).
This residual scaling method is included in the \textsc{Aips} task \textsc{Imager}
and was used recently, e.g., by \citet{Walter:2008mw} and \citet{Ianjamasimanana:2017aa}.

In principle, this method can be applied on a pixel-by-pixel basis, as long as the emissions
are significantly detected across the map. In practice, this is not always the case.
In addition, different spillovers of the beam patterns of $\Omega_{\rm syn}$ and $\Omega_{\rm conv}$
degrade the accuracy of the calculation on a pixel-by-pixel basis.
Hence, \textsc{Tp2vis} integrates the part of map with significant detections
and calculates a single beam ratio with eq. (\ref{eq:beamratio}).
Currently, \textsc{Tp2vis} uses a single ratio even for mosaics,
which may be modified in the future as the synthesized beam may vary among mosaic fields.

One may wonder why \cite{Jorsater:1995aa} could apply this technique to pure interferometer
data without single-dish. In that case $\Omega_{\rm syn}=W(0,0)=0$, and
the denominator of the left-hand side of eq. (\ref{eq:beamratio}) is apparently zero.
The answer lies in the fact that eq. (\ref{eq:beamratio}) is the ratio of surface brightnesses
in units of Jy/beam (see the right-hand side).
Both the numerator (i.e., flux) and denominator (beam) of $D-R$ are zero,
but their ratio, and hence the surface brightness, has a finite value.
Hence, this method is applicable to pure interferometer data as well.

\section{Imaging Tests}\label{sec:skymodel}

We test \textsc{Tp2vis} using model images.
With their known emission distributions, the model images allow us to evaluate
the accuracy of image deconvolution with \textsc{Tp2vis}.
We use \textsc{Casa} version 5.4.0-68 and employ the \textsc{Clean} task for imaging.
Examples with actual ALMA data are found in the \textsc{Github} repository of
\textsc{Tp2vis}, but not in this paper, as they will evolve with future updates of \textsc{Casa}.

\subsection{Setup}

\subsubsection{Model images}\label{sec:modimages}

Two model images are used for tests (Figure \ref{fig:skymodel}).
One is a single Gaussian emission distribution with a standard deviation
of 1500 pixels in width (i.e., FWHM of $\sim 53\arcsec$ in the definition described below),
and another is a pseudo sky image
that mimics small to large structures within giant molecular clouds (GMCs),
i.e., dense cores and extended emission (see Appendix \ref{sec:map4sim}).
The current \textsc{Clean}/\textsc{Tclean} tasks tend to cause flux divergence
near the edge of a mosaic field of view.
As a workaround, the emission around the edge is tapered off outward so that
it reaches zero at around the $\sim80\%$ level of the maximum primary beam coverage (see Section \ref{sec:extra}).

Both images have the same dimensions of $16,384^2$ and are configured to
the same setup, i.e., an image center coordinate of (RA, DEC) = (12h00m00s, -35d00m00s),
a cell size of $0.015\arcsec$ (hence the image size of $4.096\arcmin$),
and a frequency of 115 GHz with a bandwidth of 2 GHz (a single channel).
We run the \textsc{Simobserve} task to convert the images into the CASA format with these parameters.
The $0.015\arcsec$ cell size is chosen to include the main structures in a 38-pointing mosaic (Section \ref{sec:vis12m07m})
and corresponds to baselines of $\sim 36$ km when observed at 115 GHz.
Therefore,  the images can be properly sampled by all the ALMA configurations including
the one with the longest baseline ($\sim$16 km).
The flux scale is arbitrary, and only the relative flux with respect to the peak is important in the tests.
The peak flux is set to 1 for Figure \ref{fig:skymodel}.

\subsubsection{Visibilities for 12m and 7m Arrays}\label{sec:vis12m07m}

We employ a 38-pointing mosaic pattern with a $0.4\arcmin$ spacing.
This is a slightly finer spacing than the Nyquist sampling of the primary beam of the 12m array.
The coordinates of the 38 mosaic pointings are determined by \textsc{Simobserve}.
The same coordinates are used for the 12m, 7m, and TP arrays,
and we explicitly set the $uv$ coverage to be exactly the same for all the 38 pointings
so that the point spread function (PSF) shape does not vary across the mosaic field
except near the edge (see Section \ref{sec:extra}).

\begin{figure}
\epsscale{1.15}
\plotone{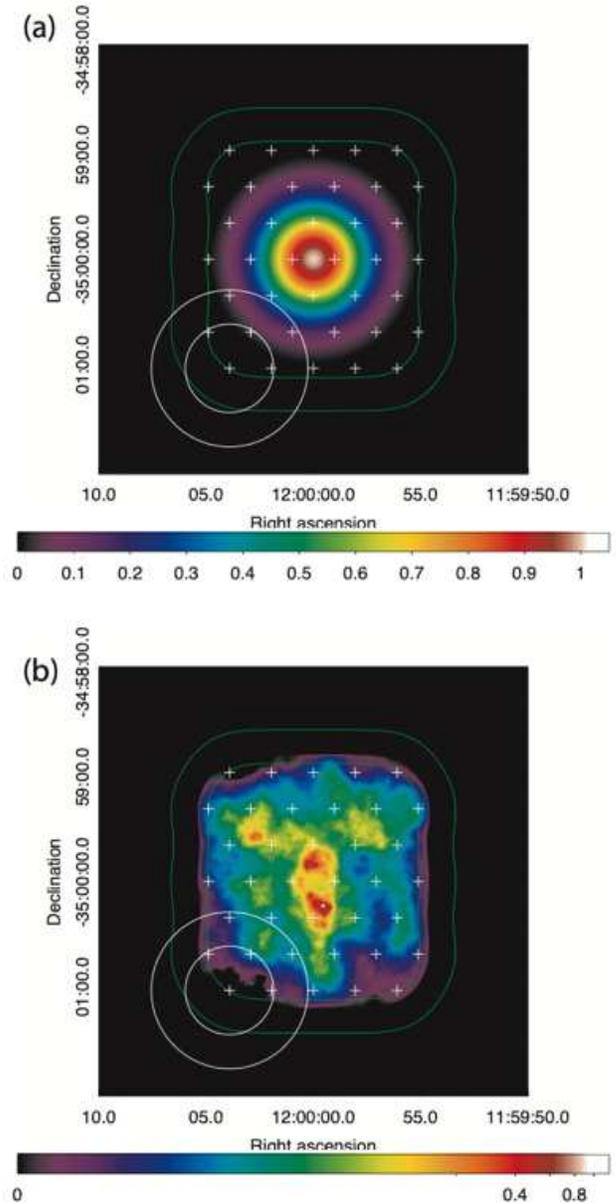}
\caption{
Model images with a dimension of $16,384^2$, corresponding to
$4.096\arcmin^2$ with the adopted cell size of $0.015\arcsec$.
The peak flux is scaled to 1.
(a) Gaussian model. A single gaussian at the center of the field of view
with a standard deviation of 1,500 pixels (i.e., FWHM$\sim 53\arcsec$).
(b) Pseudo sky model. It mimics small to large structures in GMCs (see Appendix \ref{sec:map4sim}).
The 38 mosaic pointings are plotted with crosses.
The 12m and 7m beam sizes, i.e., FWHM=$50.6\arcsec$ and $86.7\arcsec$ at 115 GHz,
are shown around the lower-left pointing for reference.
The green lines show the 40\% and 80\% levels of the peak of the 12m+7m+TP-combined
primary beam attenuation.
Note that panel (a) is in a linear scale, while panel (b) is in a log scale to show the low-level extended emission.
\label{fig:skymodel}}
\end{figure}

Figure \ref{fig:skymodel} shows the 38 pointing centers (crosses),
the 40 and 80\% levels of the  12m+7m-combined primary beam attenuation (green lines),
and the 12m and 7m beam sizes (white circles) as references.
To control the number of visibilities, we adjust total and individual integration times in \textsc{Simobserve}.
The weights of visibilities are set separately, so these times affect nothing but
the number of visibilities to be generated:
the total integration time of each pointing per array configuration
is set to 1,000 sec with a 10 sec integration.
This results in 100 visibilities per baseline.
We use one configuration for the 7m array ("aca.i.cfg") and four configurations
for the 12m array ("alma.cycle4.4.cfg", "alma.cycle4.3.cfg", "alma.cycle4.2.cfg", and "alma.cycle4.1.cfg").
The parameters of these array configurations are described in detail in the ALMA Proposer's Guide
of Cycles 4, 5, and 6.\footnote{\url{https://almascience.nrao.edu/proposing/proposers-guide}}
The tables of the antenna positions of these configurations are distributed with CASA.

\subsubsection{Visibilities for TP Array}\label{sec:visTP}

The TP array observes the sky as single-dish telescopes with a FWHM beam size of $56.7\arcsec$
at 115 GHz. TP data are delivered as a calibrated image, or cube, with a cell size of $5.6\arcsec$.
Hence, we smooth the model images with a Gaussian with a FWHM of $56.7\arcsec$
and regrid them on a $5.6\arcsec$ cell size.
We adopt the the same pointing coordinates as those of the 12m and 7m visibilities.
\textsc{Tp2vis} is run on these images and generates 5,175 TP visibilities per pointing.

\begin{figure*}
\epsscale{1.15}
\plotone{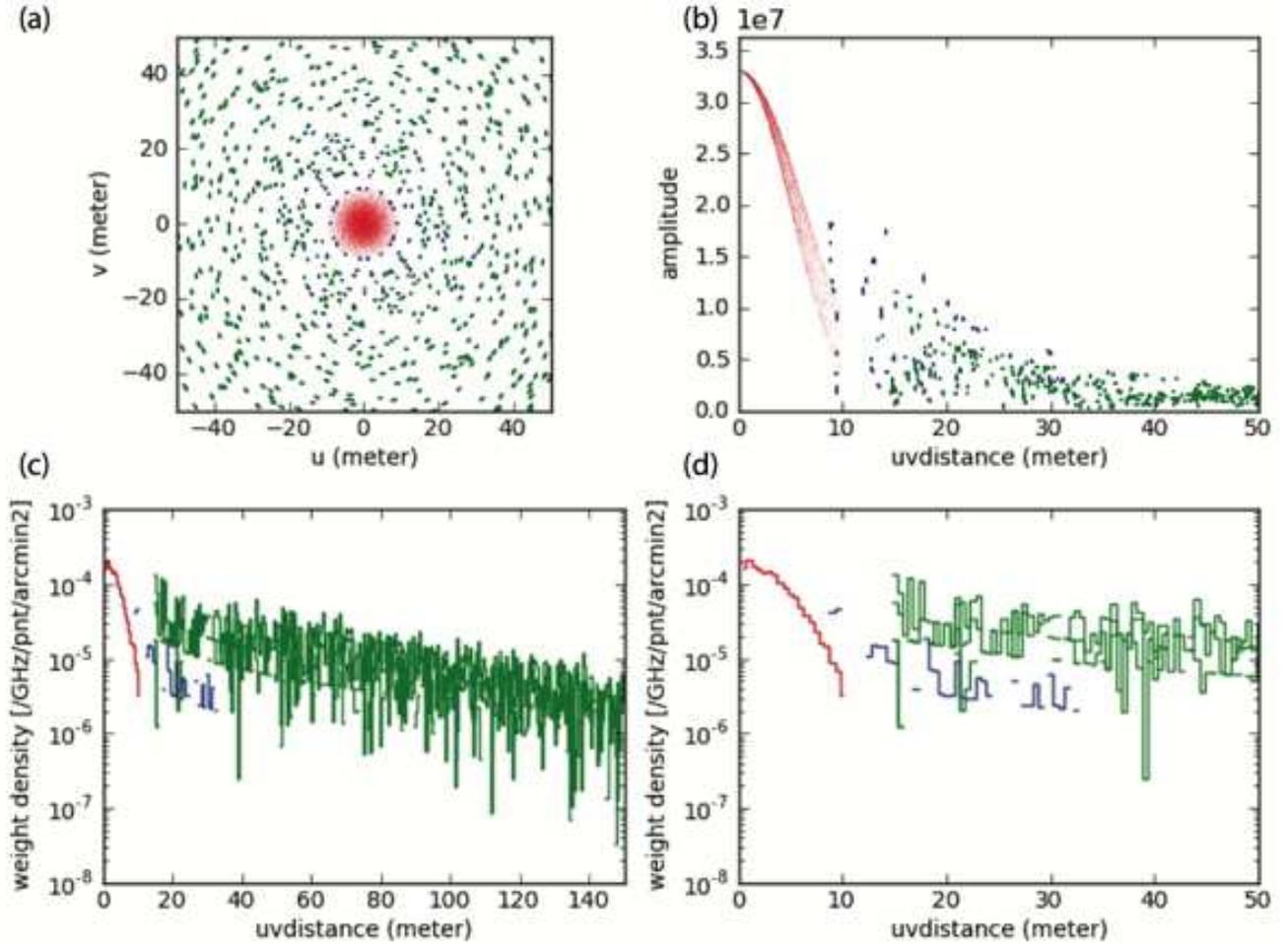}
\caption{Plot generated with \textsc{Tp2vispl} for the model data.
The red, blue, and green correspond to the TP, 7m, and 12m visibilities for the mosaic field
closest to the map center.
(a): $uv$ coverage.
(b): amplitude as a function of $uv$ distance.
(c): weight density as a function of $uv$ distance.
(d): the same as (c), but for a smaller range of $uv$ distance.
\label{fig:tp2visplot_skymodel}}
\end{figure*}

The model images do not have noise, and the visibility weights must be set arbitrarily.
We apply the sensitivity-based weight (Section \ref{sec:wtsen}) with
relative image RMS sensitivities of 1 for each of the four 12m
configurations, 4.5 for the 7m configuration, and 7 for the TP visibilities.
These relative sensitivities are chosen based on typical ALMA mosaic data we have at hand
 (e.g., of nearby galaxies, Galactic molecular clouds),
Table A-2 in the ALMA Proposer's Guide\footnote{\url{https://almascience.nrao.edu/proposing/proposers-guide\#section-57}},
and the ALMA sensitivity calculator.
The visibility and sensitivity distributions of the TP, 7m, and 12m arrays
are plotted in Figure \ref{fig:tp2visplot_skymodel}.

\subsubsection{Caveats}\label{sec:extra}

Extra efforts are needed to circumvent some of the current problems
in \textsc{Clean}/\textsc{Tclean} in \textsc{Casa}.
In general interferometer observations, the $uv$ coverages are different among mosaic pointings,
and the PSF varies across the mosaic.
The current version of \textsc{Clean}/\textsc{Tclean} (as of \textsc{Casa} 5.4)
does not take this variation into account and uses a single PSF for all mosaic pointings in its minor clean cycles.
This simplistic PSF deviates from the true PSF that reflects the $uv$ coverage at each position.
The deviation is larger for a larger mosaic, which often leads to an artificial flux divergence.
To mitigate this problem, we explicitly set the $uv$ coverage exactly the same
for all the 38 pointings so that the PSF stays the same over the most part of the entire mosaic.
This unique setup should not be a particular advantage for \textsc{Tp2vis},
since only the local PSF at the position of each emission affects clean results --
the local PSF should be consistent with the local $uv$ coverage at that position,
but it does not matter whether the PSF is the same at other parts of the mosaic.

However, we note that even with the unique $uv$ coverage for all pointings,
the edge of a mosaic suffers from the PSF variation
since it could be covered significantly by one array (e.g., the 7m, TP arrays),
but not by the others (e.g., the 12m array), due to their different primary beam sizes.
The PSF shape should be modified accordingly in such regions, as only a subset
of the arrays contributes to the $uv$ coverage.\footnote{Such a spatially-variable PSF
was implemented in the previous generation software \textsc{Miriad}, which has been
successfully applied to ALMA mosaic data \citep{Hirota:2018aa, Sawada:2018aa}.}
There is a tendency for current \textsc{Clean}/\textsc{Tclean} results to show
a flux divergence when significant emission exists near the edge.
The cause has not been entirely isolated, but the simplistic PSF likely plays a part,
since \textsc{Clean} converges when each of the 12m, 7m, and TP data are imaged separately.
Therefore, for our tests, we taper off the emission near the edge of the model images.
For the Gaussian model we subtract a small constant flux from the whole image so that the flux
decreases to zero at around the 80\% level of the peak of the 12m+7m+TP combined primary beam.
The flux outside is set to zero (Figure \ref{fig:skymodel}a).
For the pseudo sky image we first make a combined primary beam pattern image, $P$, with \textsc{Clean}.
We then multiply the original image by ``$(P^2-0.5)$", so that the flux becomes zero at around
the $\sim70-80\%$ level of the peak of the 12m+7m+TP combined primary beam. The flux outside is set to zero  (Figure \ref{fig:skymodel}b).

The \textsc{Tclean} task also suffers from the same problem and appears to have another problem
in calculation of the PSF.
These issues in \textsc{Tclean} are far beyond the scope of this paper,
but their effects and magnitudes should be characterized in the future.
Here we adopt \textsc{Clean} as it currently appears somewhat more stable than  \textsc{Tclean}
for mosaic imaging.

\subsection{Deconvolution and Results}\label{sec:casaclean}

We use the  \textsc{Clean} task for deconvolution.
The termination of  \textsc{Clean}  iterations is typically controlled by a threshold from the RMS noise,
or by the number of iterations.
The model images do not have noise, and hence we use the number of iterations
for termination and set it to $2\times10^5$.
The gain is set to 0.05. The region for clean is restricted to the area
above $80\%$ of the peak of the 12m+7m+TP combined primary beam.
We use the Briggs weighting with a ``robust" parameter of +0.5.
It results in a PSF of $1.59\arcsec \times 1.46\arcsec$ with a position angle of $87.8\deg$
for both test images.

\begin{figure*}
\epsscale{1.0}
\plotone{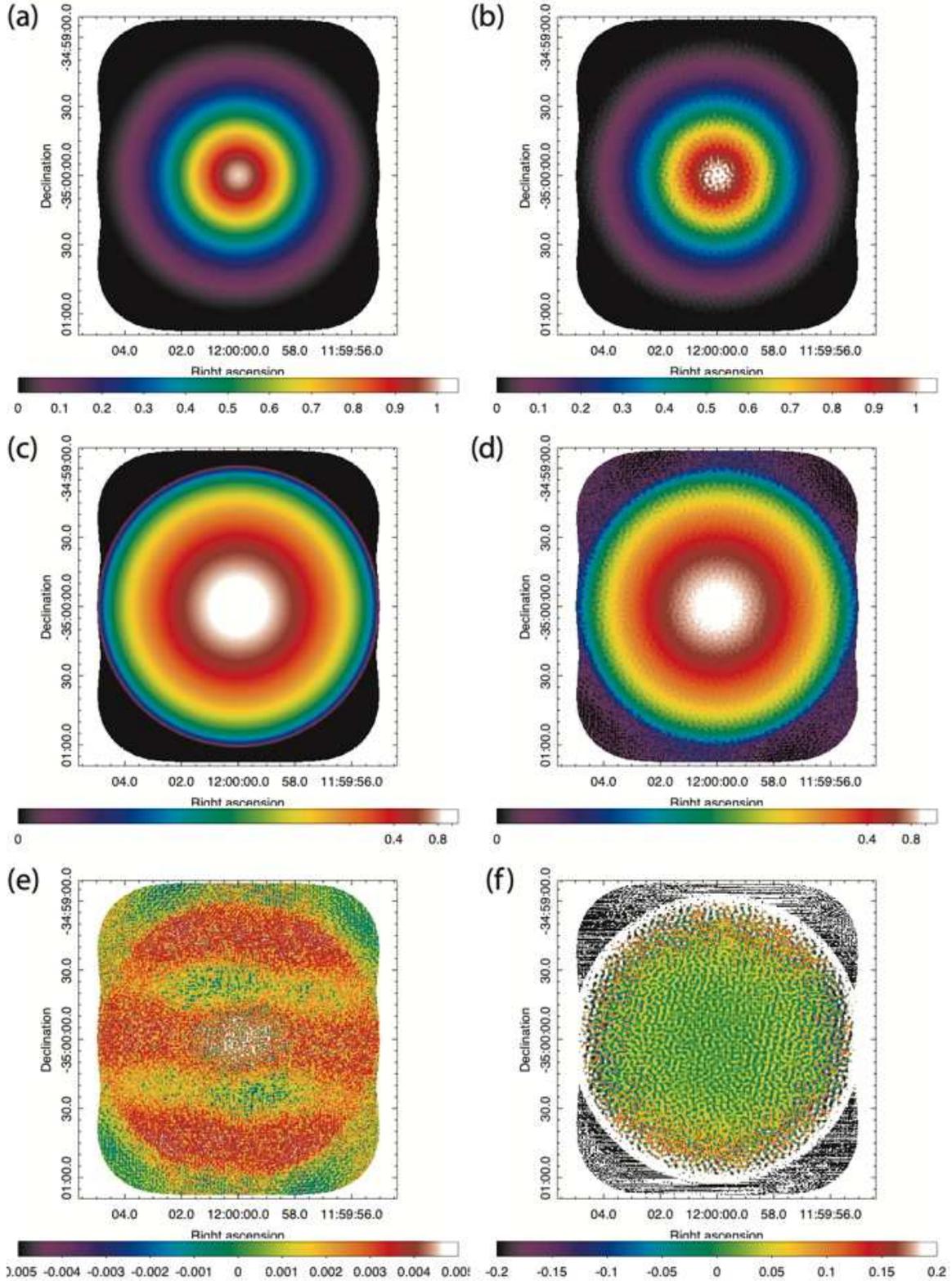}
\caption{Results of joint deconvolution of TP, 7m, and 12m visibilities for the Gaussian image.
The \textsc{Clean} task is used with niter=200,000 and gain=0.05.
The PSF size is $1.59\arcsec \times 1.46\arcsec$ with a position angle of $87.8\deg$.
(a) Model map smoothed with the PSF.
(b) Cleaned map with 12m, 7m, and TP visibilities.
(c), (d) Same as (a), (b), but on a log scale.
(e) Residual map from clean. The operation of residual scaling has been applied.
(f) Accuracy map. The deviation from the true flux is within $\sim$5\% for the most part,
and hence the quality of the cleaned map is very high.
This map can be translated to the recovered flux ratio map by adding 1 across the image.
\label{fig:gauss1500clean}}
\end{figure*}

\begin{figure*}
\epsscale{1.0}
\plotone{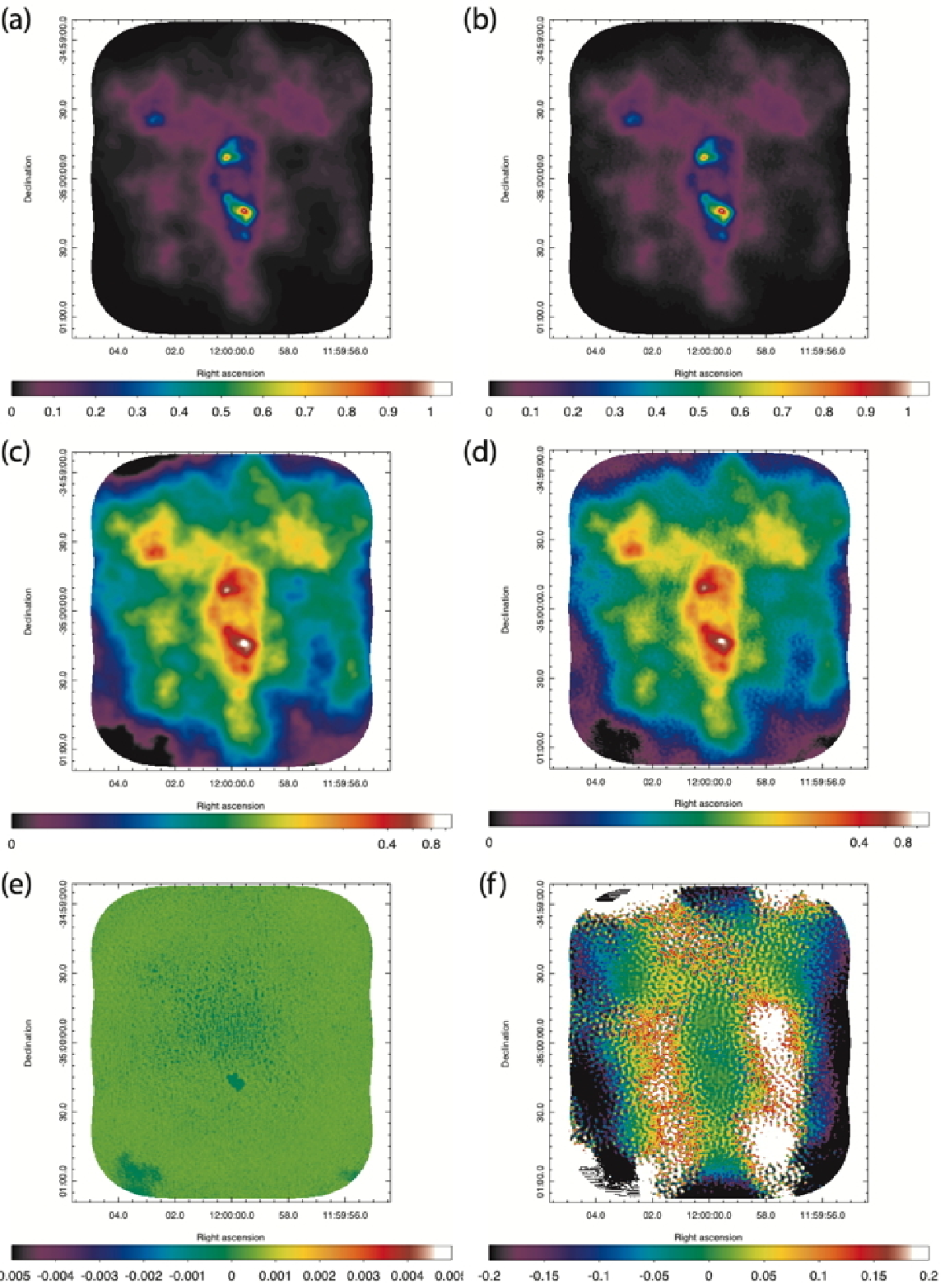}
\caption{Same as Figure \ref{fig:gauss1500clean}, but for the pseudo sky image.
\label{fig:skymodelclean}}
\end{figure*}

Figures \ref{fig:gauss1500clean} and \ref{fig:skymodelclean} compare the model and cleaned images:
(a) and (c) the original model maps smoothed with the PSF,
(b) and (d) the cleaned maps with 12m, 7m, and TP visibilities,
(e) the residual maps after the residual rescaling, and
(f) the accuracy maps.
Only relative fluxes are important for comparisons, so the fluxes of all images are normalized by
a single constant so that the peak of panel (a) becomes equal to unity.
Figures \ref{fig:gauss1500clean} and \ref{fig:skymodelclean} are normalized separately.
The accuracy maps show deviations from the input models and are calculated as
\begin{equation}
\text{Accuracy} = \frac{(\text{Cleaned image} )- (\text{Smoothed model})}{(\text{Smoothed model})}.
\end{equation}
With this definition an accuracy map plus 1 makes a recovered flux ratio map at the $1.59\arcsec \times 1.46\arcsec$ resolution.
The clean is applied over nearly three orders of magnitude in flux as the residuals after clean are $\lesssim0.3\%$ of the peak flux (panel e).
Over this wide dynamic range, the original flux is reproduced with errors mostly less than $\sim 5\%$ (panel f),
though some very diffuse regions show errors as high as $\gtrsim$20\%.

The purpose of this paper is to introduce the new \textsc{Tp2vis} package.
Hence, these tests are performed with simple noise-free images.
Further imaging simulations with more realistic setups would help evaluating systematic errors
in actual ALMA observations.

\section{Conclusions}

We presented the \textsc{Casa}-based package ``Total Power Map to Visibilities (\textsc{Tp2vis})",
which converts a TP map into visibilities.
The TP visibilities, along with 12m and 7m visibilities, can be jointly-deconvolved with standard deconvolvers (e.g., \textsc{Clean}).
The \textsc{Tp2vis} package includes four functions: \textsc{Tp2vis}, \textsc{Tp2viswt}, \textsc{Tp2vispl}, and \textsc{Tp2vistweak}. 
The \textsc{Tp2vis} function generates a Gaussian visibility distribution, whose inversion, i.e., the synthesized beam, reproduces
the primary beam of TP telescopes with natural weighting. The sensitivity-based weighing scheme is adopted as a default.
The \textsc{Tp2viswt} function can manipulate the weights.
The \textsc{Tp2vispl} function visualizes the TP weights with respect to those of 12m and 7m array data.
After the joint deconvolution, the residual map should be re-scaled with the \textsc{Tp2vistweak} function.
This rescaling is required due to the typical disparity between the areas of the synthesized beam and convolution beam.

The capability of \textsc{Tp2vis} was demonstrated using model maps.
The joint deconvolution of 12m, 7m, and TP visibilities appeared very successful and reproduced
the true emission distributions typically within 5\% errors.
The quality of images, of course, depends not only on the presence/absence of TP visibilities,
but also on the distribution and contrast of the emission. 
Tests in wider parameter spaces with different deconvolvers are beyond the scope of this paper.
Users are encouraged to do their own tests with model images that are similar to their objects of interest.
We should note that our test was done under the idealized condition of no noise,
but the 5\% accuracy over the dynamic range of $\gtrsim2$ orders of magnitude is quite encouraging.
A manual of usage is presented in the \textsc{Github} repository (\url{https://github.com/tp2vis/distribute}).

\acknowledgments

We thank the anonymous referee for useful comments,
the NRAO staff, in particular, Remy Indebetouw, Kumar Golap, Jennifer Donovan Meyer,
Crystal Brogan, and John Carpenter for their help. We also thank Kazuki Tokuda, Fumi Egusa,
Manuel Fern\'{a}ndez, and Mercedes Vazzano
for feedback on an early version of \textsc{Tp2vis} and Jim Barrett for comments on drafts.
We acknowledge a research grant from the ALMA Cycle 4 Development Study program at NRAO.
JK acknowledges supports from the National Astronomical Observatory of Japan and the Joint ALMA Observatory
during his stay on sabbatical.

\vspace{5mm}

\clearpage
\appendix

\section{GMC-like Model Map for Testing Interferometry Imaging}\label{sec:map4sim}

For tests of interferometer imaging procedures, we produce images of density/emission distributions
similar to those in giant molecular clouds (GMCs).
The method described below is similar to the one used for generating initial conditions for cosmological
simulations of large scale structures or galaxy formation, and to the one presented by \citet{Dubinski:1995lr}
for molecular cloud-like density fields. We generate a gaussian random field that follows
a density power spectrum (Section \ref{sec:densityfield}) and manipulate it to enhance the contrast (Section \ref{sec:enhance}).

\subsection{Model Density Field}\label{sec:densityfield}

We start from a representation of a density field with the density power spectrum
$P(k) \propto \left<\widetilde{\delta}^2\right> \propto k^{-n}$.
We assume a Gaussian random field (i.e., Fourier modes are uncorrelated) and
the probability distribution function of its amplitude is drawn from a Gaussian distribution.

\subsubsection{Definitions}
Using the density at position $\bold{x}$, $\rho(\bold{x})$, and average density, $\rho_0$, the density contrast is defined as
\begin{equation}
\delta(\bold{x}) \equiv \frac{\rho(\bold{x})-\rho_0}{\rho_0}.
\end{equation}

For the discrete Fourier transformation, we set the grid coordinates
\begin{equation}
\bold{x}=
D\left(
\begin{array}{c}
	l/L \\
	m/M \\
	n/N
\end{array}
\right),
\bold{k}=
\frac{1}{D}\left(
\begin{array}{c}
	u \\
	v \\
	w
\end{array}
\right),
\end{equation}
where $(l, m, n)$ and $(u, v, w)$ are integers, and $(L, M, N)$ are
the numbers of grid points in $x$, $y$, and $z$ directions, 
and $D$ is the size of the image, respectively.
We adopt the definition of the discrete Fourier transformation:
\begin{eqnarray}
\delta(l, m, n) &=& \sum_{u, v, w} \widetilde{\delta}(u, v, w) e^{+2\pi i \left(\frac{ul}{L} + \frac{vm}{M} + \frac{wn}{N} \right)}\left[ \Delta u \Delta v \Delta w\right], \label{eq:fourier}
\\
\widetilde{\delta}(u, v, w) &=& \frac{1}{LMN} \sum_{l, m, n} \delta(l, m, n) e^{-2\pi i \left(\frac{ul}{L} + \frac{vm}{M} + \frac{wn}{N} \right)}\left[ \Delta l \Delta m \Delta n\right].\label{eq:invfourier}
\end{eqnarray}
This definition is adopted in most numerical packages (e.g., IDL, Python Numpy package, etc) and is convenient for implementation.
The last terms, $\Delta u \Delta v \Delta w$ and $ \Delta l \Delta m \Delta n$ (both = 1), are written explicitly
because they need an evaluation later.
Since the density field, $\delta(l, m, n)$, is a field of real numbers, each complex Fourier component has to satisfy
\begin{equation}
\widetilde{\delta}(u, v, w) = \widetilde{\delta}^{\ast}(L-u, M-v, N-w). \label{eq:ftsym}
\end{equation}

\subsubsection{Power Spectrum and Normalization}\label{sec:pspec}

The density power spectrum is often defined in the form of a power law $P(k)= C k^{-n}$
with a normalization constant of $C$.
To avoid the singularity at $k=0$ the power spectrum is redefined here as
\begin{equation}
P(k) = \frac{C}{(k^2+k_{\rm min}^2)^{n/2}},\label{eq:pspec}
\end{equation}
where $k_{\rm min}$ is the minimum wavenumber ($k_{\rm min} = 1/D$).

The normalization constant $C$ is calculated from the variance of the density contrast, $\sigma^2$ (input parameter):
\begin{equation}
\sigma^2 = \sum_{k_u, k_v, k_w} P(k) [\Delta k_u \Delta k_v \Delta k_w] = C \sum_{k_u, k_v, k_w}  \frac{1}{(k^2+k_{\rm min}^2)^{n/2}}  [\Delta k_u \Delta k_v \Delta k_w]. \label{eq:pspec_sigv}
\end{equation}
For the purpose of analytical evaluation, we can also obtain
\begin{eqnarray}
\sigma^2 &=& \sum_{k_u, k_v, k_w} P(k) \Delta k_u \Delta k_v \Delta k_w \approx \int_0^{\infty} 4 \pi k^2 P_v(k) dk \\
 &=& 4\pi C k_{\rm min}^{3-n} \int_0^{\infty} \frac{x^2}{(x^2+1)^{n/2}}dx = 4 \pi C k_{\rm min}^{3-n} \frac{\sqrt{\pi} \Gamma \left(\frac{n-3}{2}\right)}{4 \Gamma \left(\frac{n}{2}\right)},
\end{eqnarray}
where the last equation is valid when $n>3$ (the integral diverges when $n\leq3$).
$\Gamma$ is the gamma function.
[Note $\sigma^2 \propto k_{\rm min}^{3-n} \propto D^{n-3}$,
and the size-line width relation of GMCs indicates $n\sim 4$.]
For a given set of $n$ ($>3$) and $\sigma$, the normalization coefficient of the power spectrum is
\begin{equation}
C = \frac{\sigma^2}{4 \pi} k_{\rm min}^{n-3} \frac{4 \Gamma \left(\frac{n}{2}\right)}{\sqrt{\pi} \Gamma \left(\frac{n-3}{2}\right)}.
\end{equation}

The above prescription generates a 3-dimensional data cube and involves a Fourier transformation in 3-d.
We can generate a 2-d image and apply a Fourier transformation in 2-d by using
\begin{eqnarray}
\delta^{\rm 2D}(l, m) &\equiv& \delta(l, m, n=0) \\
&=& \sum_{u,v} \left[ \sum_{w}\widetilde{\delta}(u, v, w) \right] e^{+2\pi i \left(\frac{ul}{L} + \frac{vm}{M} \right)} 
= \sum_{u,v} \widetilde{\delta}^{\rm 2D}(u, v) e^{+2\pi i \left(\frac{ul}{L} + \frac{vm}{M} \right)},
\end{eqnarray}
where
\begin{equation}
\widetilde{\delta}^{\rm 2D}(u, v) = \sum_{w}\widetilde{\delta}(u, v, w).
\end{equation}

\subsubsection{Realizations}\label{sec:realization}

The power spectrum $P(k)$ and the fluctuation of density contrast $< \widetilde{\delta}^2(k)>$ are related as follows.
\begin{eqnarray}
\sigma^2 &=& \frac{1}{LMN} \sum_{l, m, n} \delta^2(l,m,n) \left[ \Delta l \Delta m \Delta n \right] \\
&=&  \frac{1}{LMN}\sum_{u,v,w} \sum_{u',v',w'}  \widetilde{\delta}(u,v,w) \widetilde{\delta^{*}}(u',v',w')  \left\{ \sum_{l, m, n} e^{+2\pi i \left(\frac{(u-u')l}{L} + \frac{(v-v')m}{M} + \frac{(w-w')n}{N} \right)} \right\} [...][...][...]\\
&=&  \sum_{u,v,w} \widetilde{\delta}(u,v,w) \widetilde{\delta^{*}}(u,v,w) \left[ \Delta u \Delta v \Delta w \right] \label{eq:psdelta} \\
&=&  \sum_{k_u,k_v,k_w} P(k) \left[\Delta k_u \Delta k_v \Delta k_w \right]. \label{eq:pspower}
\end{eqnarray}
The $\{...\}$ term is equal to $\left\{ L \delta^K_{u,u'} \cdot M \delta^{K}_{v,v'} \cdot N \delta^{K}_{w,w'}\right\}$, where $\delta^K$ is the Kronecker delta.
The bracket terms at the end, $\left[ \Delta l \Delta m \Delta n \right]$, $\left[ \Delta u \Delta v \Delta w \right]$, etc, are expressed as $[...]$
to save space. From equations (\ref{eq:psdelta}) and (\ref{eq:pspower}),
\begin{equation}
\widetilde{\delta}(u,v,w) \widetilde{\delta^{*}}(u,v,w) \left[ \Delta u \Delta v \Delta w \right]  = P(k) \left[\Delta k_u \Delta k_v \Delta k_w \right]
\end{equation}
or
\begin{equation}
\left< \widetilde{\delta}^2(k)\right> = P(k) d^3\bold{k} = P(k) / D^3.
\end{equation}
Note that $\left[ \Delta u \Delta v \Delta w \right]=1$ and $\left[\Delta k_u \Delta k_v \Delta k_w \right]=1/D^3$.

The Fourier components $\widetilde{\delta}(k)$s are sampled using $< \widetilde{\delta}^2(k)>$ as the variance. A random number
is generated for each $\bold{k}$ using a Gaussian distribution with a standard deviation of
$\sqrt{< \widetilde{\delta}^2(k)>} = \sqrt{P(k)d^3\bold{k}}= \sqrt{P(k)/D^3}$.
The phase of the complex number $\widetilde{\delta}(k)$ is also drawn from a uniform distribution in $[0, 2\pi)$,
and the symmetry condition, eq. (\ref{eq:ftsym}), is taken into account.
Once all the Fourier components are generated, they are Fourier transformed to the image plane using equation (\ref{eq:fourier}).
We could also draw the random numbers without the symmetry condition and take the real or imaginary part
after the Fourier transformation. In this case, the amplitude should be multiplied by $\sqrt{2}$.

\begin{figure*}
\epsscale{1.0}
\plotone{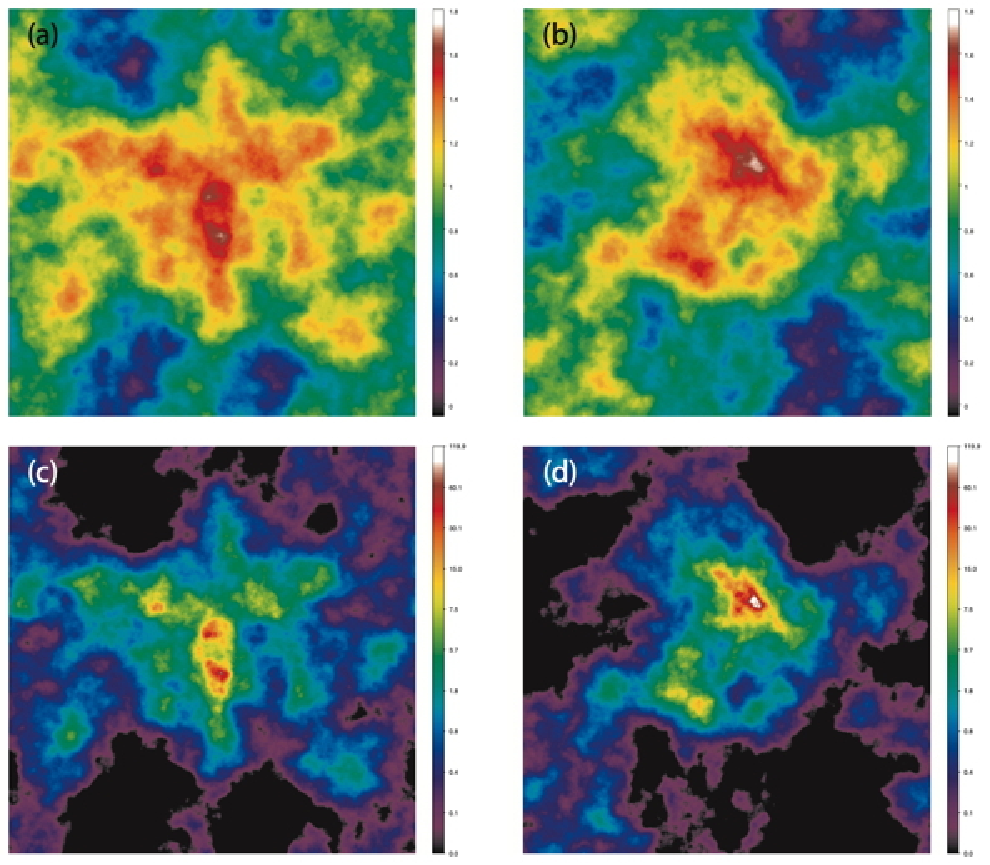}
\caption{Example model images with a power spectrum of index $n=4$.
One execution generates two realizations: (a) the real part and (b) the imaginary part,
each of which has $16,384^2$ pixels.
Panels (c) and (d) are generated from (a) and (b) by enhancing the bright parts as described in the text.
Note that (a) and (b) are in a linear scale, while (c) and (d) are in a log scale.
\label{fig:realizations}}
\end{figure*}

Figures \ref{fig:realizations}a,b show the model images generated with a power spectrum index of $n=4$,
an average density fluctuation of $\sigma=0.3$, and an image size of $16,384^2$.
Both the real and imaginary images are shown.
The images have periodic boundaries, and we spatially shifted them so that significant features come near the image centers.

\subsection{Enhance Density Contrasts}\label{sec:enhance}

Figure \ref{fig:realizations}a,b qualitatively resemble some parts of GMCs.
However, by construction, they don't have ``dense cores", whose environs are often observed with ALMA.
We could develop dense cores by running hydrodynamics simulations with gravity, starting from
these images as the initial condition, but this is obviously beyond the scope of this paper.
Here, to mimic their high brightness, we simply convert the flux $f^{\rm old}$ (or density $\rho(\bold{x})$) into new flux $f^{\rm new}$ 
using the following equation:
\begin{equation}
f^{\rm new} =
\left\{
    \begin{array}{ll}
      \exp\left[ f^{\rm old}-\beta \right] & (f^{\rm old}>\beta) \\
      0 & (otherwise).
    \end{array}
  \right.
\end{equation}
We adopt $\beta=0.7$.
This conversion is applied three times to make the images resemble GMCs with some dense cores.
The final images are displayed in Figure \ref{fig:realizations}c,d in a log scale.
They have larger dynamic ranges than the original ones, and thus, are more challenging to
reproduce with interferometer imaging.

\bibliography{tp2vis.bbl}

\end{document}